\documentclass[aps,prl,twocolumn,showpacs]{revtex4}
\usepackage{epsfig}
\usepackage{graphicx}

\begin{document}

\DeclareGraphicsExtensions{.eps,.EPS,.pdf}

\title{Competition between Bose Einstein Condensation and spin dynamics}
\author{B. Naylor$^{1,2}$, M. Brewczyk$^3$, M. Gajda$^4$, O. Gorceix$^{1,2}$, E. Mar\'echal$^{1,2}$, L. Vernac$^{1,2}$, B. Laburthe-Tolra$^{1,2}$}
\affiliation{1 Universit\'e Paris 13, Sorbonne Paris Cit\'e, Laboratoire de Physique des Lasers, F-93430
Villetaneuse, France; 2 CNRS, UMR 7538, LPL, F-93430 Villetaneuse, France; 3 Wydzia\l{ }Fizyki, Uniwersytet w Bia\l ymstoku, ul. K. Cio\l kowskiego 1L, 15-245 Bia\l ystok, Poland; 4 Institute of Physics, Polish Academy of Sciences, Aleja Lotnik\'ow 32/46,
 02-0668 Warszawa, Poland}

\begin{abstract}
We study the impact of spin-exchange collisions on the dynamics of Bose-Einstein condensation, by rapidly cooling a chromium multi-component Bose gas. Despite relatively strong spin-dependent interactions, the critical temperature for Bose-Einstein condensation is reached before the spin-degrees of freedom fully thermalize. The increase in density due to Bose-Einstein condensation then triggers spin dynamics, hampering the formation of condensates in spin excited states. Small metastable spinor condensates are nevertheless produced, and manifest strong spin fluctuations.
\end{abstract}
\date{\today}
\maketitle

Dilute quantum gases are especially suited for the investigation of non-equilibrium dynamics in closed or open quantum systems, for example associated to the physics of thermalization \cite{kaufman2016}, prethermalization \cite{gring2012}, or localization \cite{choi2016}. In particular, they provide a platform to study the kinetics of Bose-Einstein condensation. Soon after the first Bose-Einstein condensates (BECs) were obtained, it was for example possible to investigate how the BEC nucleates \cite{shvarchuk,miesner1998}. More recently, experiments performing a temperature quench below the superfluid transition investigated the dynamics of spontaneous symmetry breaking
\cite{navon2015} and revealed the production of long-lived topological defects \cite{chomaz2015}. The aim of this work is to extend the dynamical studies of Bose-Einstein condensation to the case of a multi-component Bose gas, in order to establish the mechanisms to reach both superfluid and magnetic orders. While these orders are intrinsically connected due to Bose stimulation \cite{spinor,Ueda_Review} (which contrasts with the case of Fermi fluids \cite{Coleman}), it was predicted that strong spin-dependent interactions induce spin ordering at a finite temperature \textit{above} the BEC transition \cite{SpinorOrder}.

We find that the dynamics of Bose-Einstein condensation is drastically modified due to spin-changing collisions arising from relatively strong spin-dependent interactions. Thermalization of the spin degrees of freedom is influenced by the occurrence of BEC, and in turns influences which multi-component BECs can be produced. Our experiment also demonstrates the difficulty to thermalize the spin degrees of freedom, which has a strong impact on the spin distribution of the BECs, and on their lifetime. This is of particular relevance for large spin atoms, and most notably for strongly magnetic atoms such as Cr \cite{griesmaier2005,beaufils2008}, Er \cite{aikawa}, and Dy \cite{lu} for which dipolar relaxation strongly limits the lifetime of multicomponent gases \cite{pasquiou2010}.

\begin{figure}
\centering
\includegraphics[width= 8 cm]{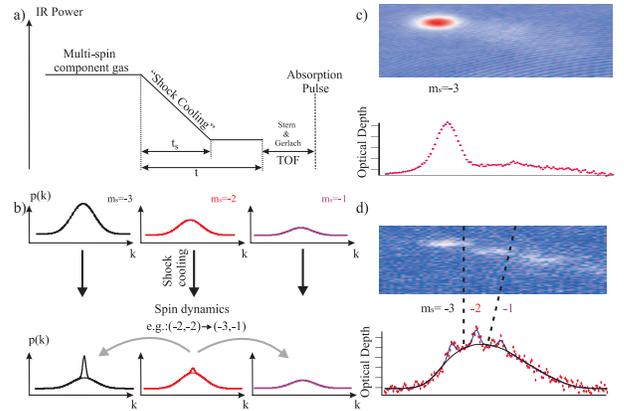}
\caption{a) Experimental sequence, showing the reduction in the ODT intensity in a duration $t_{S}$. An absorption image is taken after a time $t$ and Stern-Gerlach separation. b) Simple cartoon of the evolution of the momentum  distributions ($p(k)$) of atoms in the three lowest spin excited states, illustrating the difficulty of achieving BEC (peak on top of the broad thermal $k$ distribution) in spin excited states due to spin dynamics. Absorption pictures showing: (c) a BEC in $m_s=-3$ and a thermal gas in other spin states for a gas initially prepared with magnetization $M=-2.5\pm$0.25; d) a small multi-component BEC for $M=-2.00\pm0.25$.}

\label{Absorption}
\end{figure}

We induce fast evaporative cooling of a multi-component s$=$3 chromium thermal cloud by lowering the depth of a spin-insensitive optical dipole trap (ODT) (see Fig \ref{Absorption}a). When the gas is only slightly depolarized, the thermal gas of the most populated, lowest energy, $m_s=-3$ state rapidly saturates (i.e. reaches the maximal number of atoms in motional excited states allowed by Bose statistics \cite{ExpSaturation}) and a BEC is produced for this spin state. Saturation is also reached for the thermal gas in the second-to-lowest energy state $m_s=-2$. However, surprisingly, this state fails to condense and the BEC remains fully magnetized. In contrast, when the experiment is performed with an initially more depolarized thermal gas, spinor (i.e. multi-component) condensates are obtained, although they remain very small (see Fig. \ref{Absorption}) and show strong spin fluctuations. Comparison with numerical simulations based on the classical field approximation (CFA) \cite{CFA1} reveals that the difficulty to obtain a multi-component  BEC is due to spin exchange collisions, which rapidly empty the condensates in spin excited states by populating spin states for which the thermal gas is not yet saturated. There is  an intriguing interplay between condensation and spin dynamics, as the large increase in density associated to BEC triggers fast spin dynamics which in turn tends to deplete the BEC in spin excited states. The observed spin fluctuations in the BEC are ascribed to a combined effect of phase fluctuations due to symmetry-breaking at the BEC transition, and spin dynamics.

To prepare an incoherent spin mixture of thermal gases, we start from a thermal gas of $2.10^4$ $^{52}$Cr atoms, at $T=1.1\times T_c=440\pm 20$nK, polarized in the Zeeman state $m_s=-3$. We adiabatically reduce the magnetic field $B$ so that the Zeeman energy is of the same order as the thermal kinetic energy. Depolarization of the cloud is driven by magnetization-changing collisions associated to dipole-dipole interactions \cite{DemagnetizationThermodynamics,DemagnetizationCooling}. We obtain a gas of longitudinal magnetization $M=-2.50\pm$0.25, with $M \equiv \sum_{i=-s}^{s}i n_i$ ($n_i$ is the relative population in Zeeman state $m_s=i$). We then reduce the trap depth by applying an approximately linear ramp to the ODT laser intensity in a time $t_{S}$. This  results in fast forced evaporative cooling of all Zeeman states (which we refer to as "shock cooling", see Fig. \ref{Absorption}a). We study spin dynamics and condensation dynamics by measuring both the spin and momentum distributions as a function of the time $t$ after the beginning of the evaporation ramp. This measurement is performed by switching off all trapping lights, and applying an average magnetic field gradient of 3.5 G.cm$^{-1}$ during a 6 ms time of flight to perform a Stern-Gerlach separation of the free-falling atoms.

\begin{figure}[t]
\centering
\includegraphics[width= 7 cm]{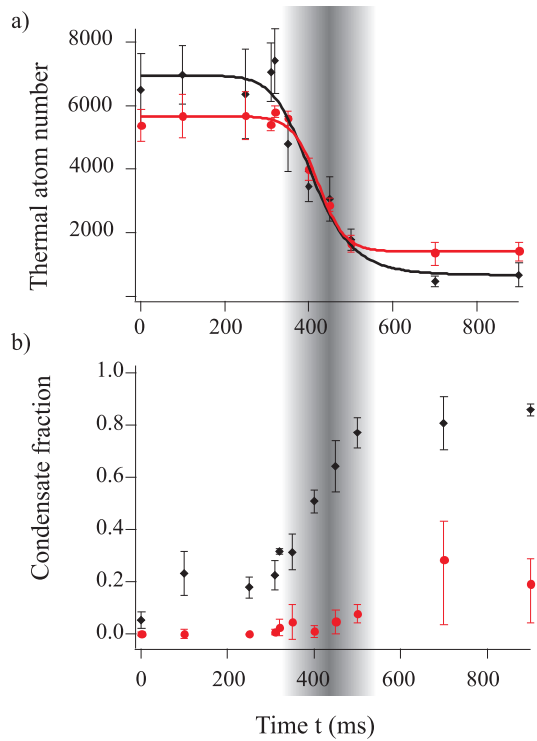}
\caption{a) Number of thermal atoms in $m_s=-3$ (black diamonds) and $m_s=-2$ (red disks) as a function of time $t$ for a shock cooling time $t_S=500$ ms. b) Corresponding condensate fractions in $m_s=-3$ (black diamonds) and $m_s=-2$ (red disks). The shaded region highlights when both $m_s=-3$ and $m_s=-2$ thermal clouds are saturated, but only $m_s=-3$ atoms condense.}
\label{Saturation}
\end{figure}

Fig.\ref{Absorption}c shows a typical absorption picture. It reveals a BEC in $m_s=-3$, and a thermal gas in spin excited states. We extract the number of thermal and condensed atoms of each spin state through bi-modal fits accounting for Bose statistics. We plot in Fig.\ref{Saturation} the thermal atom numbers as well as the condensate fractions in $m_s=(-3,-2)$ as a function of time $t$ for a shock cooling time $t_S=500$ ms. We found similar results for $t_S =$ 250 ms and $t_S =$ 1 s.

The gray area in Fig.\ref{Saturation} highlights a relatively long cooling time during which the $m_s=-3 $ and $m_s=-2$ gases hold approximately the same number of thermal atoms, and there is a BEC in the lowest state $m_s=-3$ but not in $m_s=-2$. This phenomenon is surprising, because it shows that the $m_s=-2$ component fails to undergo Bose-Einstein condensation although its thermal gas is saturated. Indeed, $m_s=-2$ and $m_s=-3$ thermal atoms have the same measured mechanical temperature (within our 5\% experimental uncertainty), experience the same trapping potential, and  both interact through the $S=6$ molecular potential with the existing $m_s=-3$ BEC. Therefore the $m_s=-2$ cloud has the same thermodynamical properties than the $m_s=-3$ thermal gas, and like this latter spin component, should condense for further cooling \cite{ExpSaturation}. However, BEC does not occur in this state, until $t \leq 700$ ms. Only for $t \geq 700$ms do we distinguish a very small BEC also in $m_s=-2$. This demonstrates that a BEC in $m_s=-2$ hardly forms, although the thermal gas is saturated and cooling proceeds.

To interpret these observations, we stress that magnetization-changing collisions occur on a larger time scale than shock cooling dynamics and can be neglected, contrarily to \cite{DemagnetizationThermodynamics}. Here, spin dynamics is almost entirely controlled by spin exchange interactions at constant magnetization driven by spin dependent contact interactions \cite{Ueda_Review}. A key point is that for an incoherent mixture the spin dynamics rate $\gamma_{i,j}^{k,l}$ for the spin changing collision $(m_s=i,m_s=j) \rightarrow (m_s=k,m_s=l)$ is set by the density of the cloud through $\gamma_{i,j}^{k,l}=n\sigma_{i,j}^{k,l} v$ with $n$ the atomic density, $v$ the average relative atomic velocity and $\sigma_{i,j}^{k,l}$ the relevant cross section within Born approximation \cite{Dynamics}. This rate is extremely sensitive to the presence of a BEC (which enhances $n$). Therefore, the emergence of a BEC in a spin-excited state should trigger faster spin dynamics. In addition to these dissipative spin-exchange processes, BEC can also trigger coherent spin oscillations due to forward scattering, with a typical rate $\Gamma_{i,j}^{k,l} = \frac{4 \pi \hbar}{m} n \sum_S a_S \langle i,j | S \rangle    \langle  S | k,l \rangle$ where the sum is on even molecular potentials $S$, with associated scattering length $a_S$. Our interpretation for the absence of a BEC in the state $m_s=-2$ is thus that a large BEC cannot form in this state because fast spin-exchange processes $(-2,-2) \rightarrow (-1,-3)$ deplete the BEC as soon as it is produced. Thus spin dynamics and condensation dynamics are strongly intertwined.

To check this interpretation, we have performed numerical simulations using the Gross-Pitaevskii (GP) equation and the classical field approximation to describe thermal states. According to CFA, the GP equation determines the evolution of the classical field which is a complex function carrying the information on both the condensed and thermal atoms \cite{CFA1,CFA2,Karpiuk10}. The initial classical field corresponds to $13.10^3$ Cr atoms at the critical temperature of about $400\,$nK and with the experimental Zeeman distribution. To describe such a sample we follow the prescription given in \cite{Karpiuk10}. Evaporative cooling is mimicked by adding a purely imaginary potential to the GP equation at the edge of the numerical grid.

\begin{figure}
\centering
\includegraphics[width=7cm]{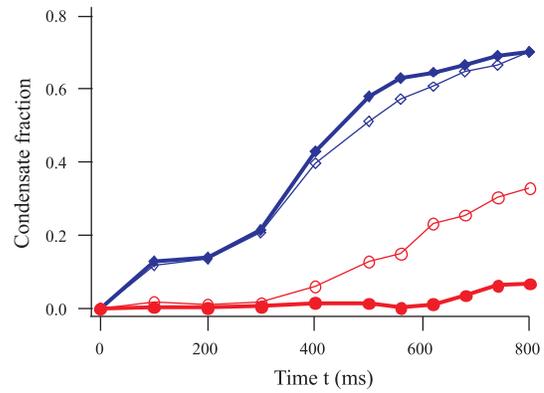}
\caption{Numerical results. Evolution of the condensate fractions for different values of $a_4$  (blue diamonds: $m_s=-3$; red circles: $m_s=-2$). Filled markers correspond to the experimental case: $a_4=64$ $a_B$ and $a_6=102.5$ $ a_B$ \cite{pasquiou2010} where $a_B$ is the Bohr radius. Empty markers correspond to simulations where $a_4$ was set equal to $a_6$ to suppress spin dynamics. A significative BEC fraction in $m_s=-2$ is then obtained.}
\label{Theorie}
\end{figure}

Our simulations confirm the existence of a saturated $m_s=-2$ gas and the absence of a condensate in this state (see Fig.\ref{Theorie}). To evaluate the impact of spin-exchange processes on the dynamics of condensation, we have reproduced these simulations assuming $a_4=a_6$. In this case, the rates associated to spin-exchange processes $(-2,-2) \rightarrow (-1,-3)$, $\gamma_{-2,-2}^{-1,-3}$ and $\Gamma_{-2,-2}^{-1,-3}$, which respectively scale as $(a_6-a_4)^2$ and $(a_6-a_4)$, both vanish. As shown in Fig. \ref{Theorie}, a BEC then forms in the spin excited state $m_s=-2$. This confirms the crucial role of spin-dependent interactions in the dynamics of Bose-Einstein condensation.

It is interesting to face our observations with the accepted scenario for the thermodynamics of non-interacting multi-component Bose gases at fixed magnetization \cite{Theory_PhaseTransition}. In this picture, a BEC polarized in the most populated state forms below a first critical temperature; all the other thermal spin states saturate \textit{simultaneously} and condense below a second critical temperature \cite{Theory_PhaseTransition}. In our situation, our observations indicate that the external degrees of freedom have reached an equilibrium, at an effective temperature which we find identical for all spin components. However, although the thermal clouds of the two lowest spin components are saturated, the other thermal clouds are not saturated. This is in profound contradiction with the prediction of Bose thermodynamics, and shows that the spin degrees of freedom in our experiment remain out of equilibrium.

This lack of thermal equilibrium for the spin degrees of freedom results from the fact that spin exchange processes \textit{for the thermal gas} are slow in regards of condensation dynamics. For example, the rate of the dominant spin exchange term, averaged over density, $\gamma_1=\frac{n_0\sigma_{-2,-2}^{-1,-3}v}{2\sqrt{2}}$, with $n_0$ the peak atomic density, is typically 3 $s^{-1}$ for a thermal gas at $T_C$. A much longer timescale would therefore be necessary in order to reach spin equilibrium. This rate is slow compared to typical thermalization rates of the mechanical degrees of freedom e.g. $\gamma_2=\frac{n_0\sigma_{-2,-2}^{-2,-2}v}{2\sqrt{2}}\approx $ 40 s$^{-1}$. $\gamma_2>>\gamma_1$ insures that the mechanical degrees of freedom thermalize faster than the spin degree of freedom, and that a small $m_s=-2$ BEC can in principle be formed. However, once the $m_s=-2$ BEC is formed, the rates associated to $(-2,-2)\rightarrow(-3,-1)$ collisions rise to typically $\gamma_{1,BEC}\approx$15 s$^{-1}$ and $\Gamma_{-2,-2}^{-1,-3} \approx 100 $ s$^{-1}$ (for 500 atoms in the condensate). Spin exchange collisions then deplete the $m_s=-2$ BEC as fast as it is created and a multi-component BEC cannot be sustained due to the lack of saturation of the $m_s=-1$ thermal gas.

Under our experimental conditions, non-saturated spin-excited states thus act as a reservoir into which population may be dumped, thus preventing BEC but in the stretched state, the only collisionally stable one. The situation bears some analogies to the condensation of magnons \cite{Magnon_StamperKurn,Magnon_Condensation} and polaritons, where BEC is obtained in the lowest momentum state by collisions of higher states in the lower polariton branch \cite{Polariton_Condensation}. Like for polaritons, it is likely that spin-exchange interactions are increased by Bose-stimulation due to the pre-existing $m_s=-3$ condensate.

\begin{figure}
\centering
\includegraphics[width=8cm]{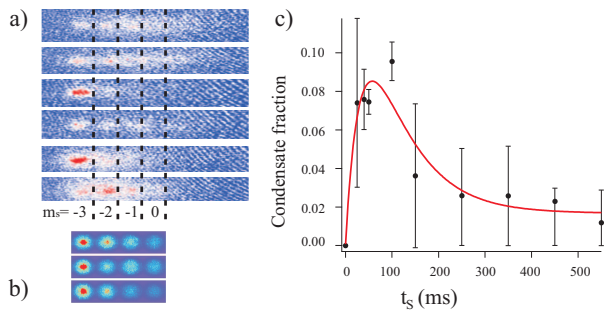}
\caption{a) Absorption images after shock cooling experiments performed with $t_S=50$ ms and an initial magnetization $M=-2\pm0.25$. A small BEC is present in the three lowest spin states i.e. $m_s=$ -3,-2, and -1. The different images illustrate the fluctuations of magnetization of the condensate fraction. b) Numerical results after evaporation, for three initial relative sets of phases. c) Total condensate fraction of the multi spin component gas for $t=t_S$ ms and $M=-2\pm0.25$ as a function of $t_S$. We observe small multi-component condensates in the three lowest energy states. The solid line guides the eye.}
\label{MultiBEC}
\end{figure}

To produce multi-component condensates during evaporation, we performed a second series of shock cooling experiments, with a lower gas magnetization $M=-2.00\pm0.25$ (where the uncertainty is associated to detection noise). The initial $m_s=-3$ thermal gas is now depolarized using a radio-frequency pulse. After decoherence of the spin components, this leads to a thermal incoherent mixture with initial fractional population in $m_s=-3$, $-2$, $-1$ and $0$ states approximately (31\%,40\%,21\%,6\%) with a relative uncertainty of 10\%. When shock cooling is performed fast compared to $\gamma_{1,BEC}$, we observe the production of very small multi-component condensates in all three lowest energy states (as illustrated in Fig \ref{Absorption} and \ref{MultiBEC}). Our numerical simulations show that spin-dynamics has again a very profound influence on the dynamics of condensation. In practice, spin excited states with $m_s>0$ are not saturated. Therefore, spin dynamics tends to populate these non saturated states and empty the condensates, which thus remain small and short-lived.

An important observation is that the spin distribution of the obtained multi-component BECs shows strong fluctuations (see Fig.\ref{MultiBEC}a) compared to the thermal fraction. We interpret this feature in the following way. Bose-condensation of the different spin-excited states introduce a spontaneous symmetry breaking as the phase of each condensate is chosen randomly. This spontaneous symmetry breaking, already observed in \cite{Magnon_StamperKurn}, can also be interpreted as the production of fragmented BECs \cite{Bigelow_Fragmentation, Gerbier_Fragmentation}. As spin-dynamics is sensitive to the relative phase between the condensates in the various spin states \cite{Chapman, Ueda_Review}, we propose that the observed spin fluctuations result from a combined effect of spin dynamics and of the spontaneous symmetry breaking.

We performed numerical simulations to test this scenario. As the CFA does not provide a direct way to provide symmetry breaking at the BEC transition, we chose to apply random relative phases to the wave-functions describing the thermal atoms in different spin components before condensation. This provides an empirical way to simulate symmetry breaking. We performed a series of numerical simulations for different sets of relative phases between the Zeeman components. We then obtained small condensates with fluctuating magnetization (see Fig.\ref{MultiBEC}b). Furthermore, we also observe that spin and condensation dynamics are also significantly modified by the applied random phases. Due to large computational time for each run, a systematic study of BEC magnetization as a function of initial phases has not yet been performed and remains to be thoroughly investigated. However, while the magnetization fluctuations obtained in the numerical simulations are typically five time smaller than the experimental measurements, these preliminary results thus support the scenario that the combined effect of spontaneous symmetry breaking and spin dynamics lead to the observed spin fluctuations.

As a conclusion, our study reveals a strong interplay between Bose condensation and spin dynamics, which is of particular relevance when spin-dependent and spin-independent interactions take place on a similar timescale (in contrast to previous studies with alkali atoms, see \cite{schmaljohann2004}). This interplay can for example result in a delay in obtaining a BEC in spin-excited states or alternatively to the production of weak metastable spinor gases which decay due to spin-exchange collisions. Our results also show that the difficulty to fully thermalize the spin degrees of freedom is a prominent effect to be taken into account for very large spin systems (such as Dy \cite{lu} and Er \cite{aikawa}), where all spin states must be saturated for a stable multi-component BEC to be produced. Finally, we point out that when a multi component BEC is dynamically produced, spontaneous symmetry breaking leads to independent phases within the BEC components which triggers spin fluctuation.

This work was supported by Conseil R\'egional d'Ile-de-France under DIM Nano-K/IFRAF, Minist\`ere de l'Enseignement Sup\'erieur et de la Recherche within CPER Contract, Universit\'e Sorbonne Paris Cit\'e (USPC), and by  the  Indo-French  Centre  for  the  Promotion  of Advanced Research-CEFIPRA. M.B. and M.G. acknowledge  support of the National Science Center (Poland) Grant No. DEC-2012/04/A/ST2/00090.


\begin{thebibliography}{199}


\bibitem{kaufman2016} Adam M. Kaufman, M. Eric Tai, Alexander Lukin, Matthew Rispoli, Robert Schittko, Philipp M. Preiss, Markus Greiner, arXiv:1603.04409 (2016)

\bibitem{gring2012} Michael Gring, Maximilian Kuhnert, Tim Langen, Takuya Kitagawa, Bernhard Rauer, Matthias Schreitl, Igor Mazets, David A. Smith, Eugene Demler, Jörg Schmiedmayer,   Science \textbf{337}, 1318 (2012)

\bibitem{choi2016}  Jae-yoon Choi, Sebastian Hild, Johannes Zeiher, Peter Schauß, Antonio Rubio-Abadal, Tarik Yefsah, Vedika Khemani, David A. Huse, Immanuel Bloch, Christian Gross,  arXiv:1604.04178 (2016)

\bibitem{shvarchuk} I. Shvarchuck, Ch. Buggle, D. S. Petrov, K. Dieckmann, M. Zielonkowski, M. Kemmann, T. G. Tiecke, W. von Klitzing, G. V. Shlyapnikov, and J. T. M. Walraven, Phys. Rev. Lett. \textbf{89}, 270404 (2002)

\bibitem{miesner1998} H.-J. Miesner et al., Science \textbf{279}, 1005 (1998).

\bibitem{navon2015} Nir Navon, Alexander L. Gaunt, Robert P. Smith, Zoran Hadzibabic, Science \textbf{347}, 167-170 (2015)

\bibitem{chomaz2015} Lauriane Chomaz, Laura Corman, Tom Bienaim\'e, R\'emi Desbuquois, Christof Weitenberg, Sylvain Nascimb\`ene, J\'erôme Beugnon, Jean Dalibard, Nature Communications \textbf{6}, 6172 (2015)

\bibitem{spinor} Tin-Lun Ho, Phys. Rev. Lett. \textbf{81}, 742 (1998), T. Ohmi, K. Machida, J. Phys. Soc. Jpn. \textbf{67},1822 (1998), D. M. Stamper-Kurn and M. Ueda, Rev. Mod. Phys. \textbf{85}, 1191 (2013)

\bibitem{Ueda_Review} Y. Kawaguchi and M. Ueda Physics Reports, Volume 520, Issue 5, 253 (2012)

\bibitem{Coleman} Piers Coleman, Introduction to many body physics, Cambridge University Press (January
 15, 2016); Maciej Lewenstein, Anna sanpera, Veronica Ahufinger, Ultracold atoms in optical lattices, Oxford University Press, 2012

\bibitem{SpinorOrder} Stefan S. Natu and Erich J. Mueller, Phys. Rev. A \textbf{84}, 053625 (2011)

\bibitem{griesmaier2005} Axel Griesmaier, Jörg Werner, Sven Hensler, Jürgen Stuhler, and Tilman Pfau, Phys. Rev. Lett. \textbf{94}, 160401 (2005)

\bibitem{beaufils2008} Q. Beaufils, R. Chicireanu, T. Zanon, B. Laburthe-Tolra, E. Mar\'echal, L. Vernac, J.-C. Keller, and O. Gorceix, Phys. Rev. A \textbf{77}, 061601(R) (2008), B. Naylor, A. Reigue, E. Mar\'echal, O. Gorceix, B. Laburthe-Tolra, and L. Vernac, Phys. Rev. A \textbf{91}, 011603(R) (2015)

\bibitem{aikawa} K. Aikawa, A. Frisch, M. Mark, S. Baier, A. Rietzler, R. Grimm, and F. Ferlaino, Phys. Rev. Lett. \textbf{108}, 210401 (2012), K. Aikawa, A. Frisch, M. Mark, S. Baier, R. Grimm, and F. Ferlaino, Phys. Rev. Lett. \textbf{112}, 010404 (2014)

\bibitem{lu} Mingwu Lu, Nathaniel Q. Burdick, Seo Ho Youn, and Benjamin L. Lev, Phys. Rev. Lett. \textbf{107}, 190401 (2011), Mingwu Lu, Nathaniel Q. Burdick, and Benjamin L. Lev, Phys. Rev. Lett. \textbf{108}, 215301 (2012)

\bibitem{pasquiou2010} B. Pasquiou, G. Bismut, Q. Beaufils, A. Crubellier, E. Mar\'echal, P. Pedri, L. Vernac, O. Gorceix, and B. Laburthe-Tolra, Phys. Rev. A \textbf{81}, 042716 (2010).

\bibitem{ExpSaturation} Tammuz, N., Smith, R.P., Campbell, R.L.D., Beattie, S., Moulder, S., Dalibard, J., and Hadzibabic, Z., Phys. Rev. Lett. \textbf{106}, 230401 (2011).

\bibitem{CFA1} K. G\'oral, M. Gajda, and K. Rz\c a\.zewski, Opt. Express {\bf 10}, 92 (2001).

\bibitem{DemagnetizationThermodynamics} B. Pasquiou, E. Mar\'echal, L. Vernac,O. Gorceix, and B. Laburthe-Tolra, Phys. Rev. Lett. \textbf{108}, 045307 (2012).

\bibitem{DemagnetizationCooling} M. Fattori \textit{et al.}, Nature Phys. \textbf{2}, 765 (2006).

\bibitem{Dynamics} Ulrich Ebling, Jasper Simon Krauser, Nick Fläschner, Klaus Sengstock, Christoph Becker, Maciej Lewenstein, and Andr\'e Eckardt
Phys. Rev. X \textbf{4}, 021011 (2014)

\bibitem{CFA2} M. Brewczyk, M. Gajda, and K. Rz\c a\.zewski, J. Phys. B {\bf 40}, R1 (2007).

\bibitem{Karpiuk10} T. Karpiuk, M. Brewczyk, M. Gajda, and K. Rz\c a\.zewski, Phys. Rev. A {\bf 81}, 013629 (2010).

\bibitem{Theory_PhaseTransition} T. Isoshima, T. Ohmi, K. Machida, J. Phys. Soc. Jpn. \textbf{69},3864 (2000).

\bibitem{Magnon_StamperKurn} Fang Fang, Ryan Olf, Shun Wu, Holger Kadau, and Dan M. Stamper-Kurn  Phys. Rev. Lett. \textbf{116}, 095301 (2016)

\bibitem{Magnon_Condensation} S. O. Demokritov, V. E. Demidov, O. Dzyapko, G. A. Melkov, A. A. Serga, B. Hillebrands and A. N. Slavin Nature \textbf{443}, 430-433 (2006)
\bibitem{Polariton_Condensation} J. Kasprzak, M. Richard, S. Kundermann, A. Baas, P. Jeambrun, J. M. J. Keeling, F. M. Marchetti, M. H. Szyman acuteska, R. Andr\'e, J. L. Staehli, V. Savona, P. B. Littlewood, B. Deveaud and Le Si Dang, Nature \textbf{443}, 409-414 (2006).



\bibitem{Bigelow_Fragmentation} C. K. Law, H. Pu, and N. P. Bigelow, Phys. Rev. Lett. \textbf{81}, 5257 (1998), Tin-Lun Ho and Sung Kit Yip, Phys. Rev. Lett. \textbf{84}, 4031 (2000)

\bibitem{Gerbier_Fragmentation} Luigi De Sarlo, Lingxuan Shao, Vincent Corre, Tilman Zibold, David Jacob, Jean Dalibard and Fabrice Gerbier New Journal of Physics \textbf{15} 113039 (2013)

\bibitem{Chapman} M.S. Chang, Q. Qin, W. Zhang, L. You, M.S. Chapman, Nature Physics \textbf{1}, 111-116 (2005)

\bibitem{schmaljohann2004} H. Schmaljohann, M. Erhard, J. Kronjäger, K. Sengstock, K. Bongs, Applied Physics B, \textbf{79}, 1001 (2004)

\end{thebibliography}
\end{document}